\documentclass[a4paper, 12pt]{article}

\usepackage{amssymb,amsmath,mathrsfs,subfigure,epsfig, color}
\usepackage{amsmath,natbib}
\usepackage[latin1]{inputenc}

\def\d{\mathrm{d}}
\def\im{\mathrm{i}}

\def \div{\mbox{div\hskip 1pt}}

\def \tr{\mbox{tr\hskip 1pt}}
\def \grad{\mbox{grad\hskip 1pt}}

\def\eps{\varepsilon}
\begin{document}

\title{Straightening Wrinkles}

\author{Michel Destrade$^{a,b}$, Ray W. Ogden$^c$, Ivonne Sgura$^d$, Luigi Vergori$^a$ \\[12pt]
$^a$School of Mathematics, Statistics \& Applied Mathematics, \\ National University of Ireland Galway, Ireland;\\[6pt]
$^b$School of Mechanical \& Materials Engineering, \\ University College Dublin, Belfield, Dublin 4, Ireland;\\[6pt]
$^c$School of Mathematics \& Statistics, University of Glasgow, UK;\\[6pt]
$^d$Dipartimento di Matematica e Fisica,\\ Universit\`a del Salento, Lecce, Italy.}

\date{}

\maketitle

\bigskip
\bigskip


\begin{abstract}

We consider the elastic deformation of a circular cylindrical sector composed of an incompressible isotropic soft solid when it is straightened into a rectangular block. 
In this process, the circumferential line elements on the original inner face of the sector are stretched while those on the original outer face are contracted. 
We investigate the geometrical and physical conditions under which the latter line elements can be contracted to the point where a \color{black} localized \color{black} incremental instability develops. We provide a robust algorithm to solve the corresponding two-point boundary value problem, which is stiff numerically. 
We illustrate the results with full \color{black} incremental \color{black} displacement fields in the case of Mooney--Rivlin materials and also perform an asymptotic analysis for thin sectors. 

\end{abstract}


\bigskip


\newpage


\section{Introduction}

\begin{figure}[ht!]\label{fig0}
\centering
\subfigure[]{\includegraphics[width=0.5\textwidth, keepaspectratio]{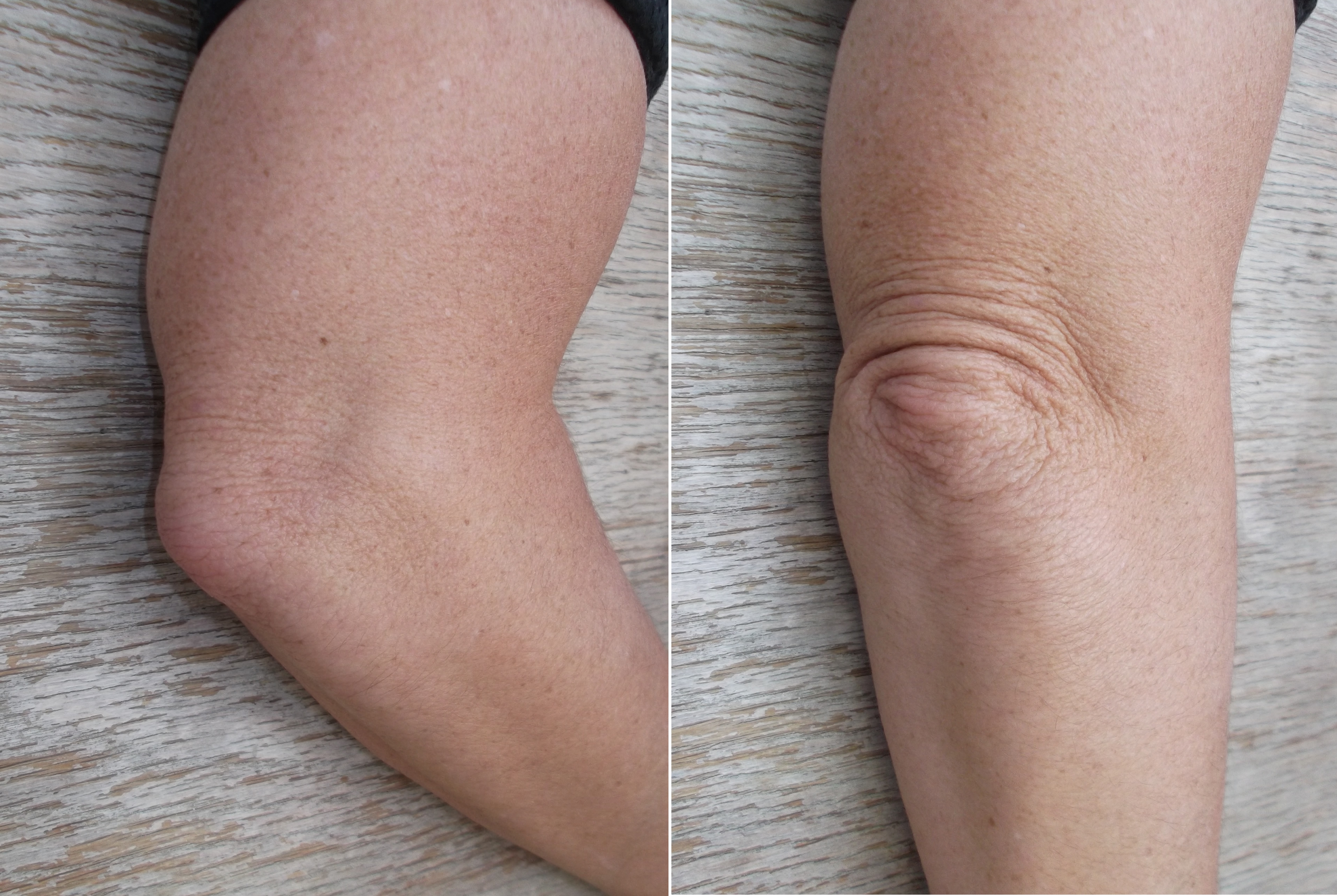}}\hspace{1cm}
	\subfigure[] {\includegraphics[width=0.4\textwidth, keepaspectratio]{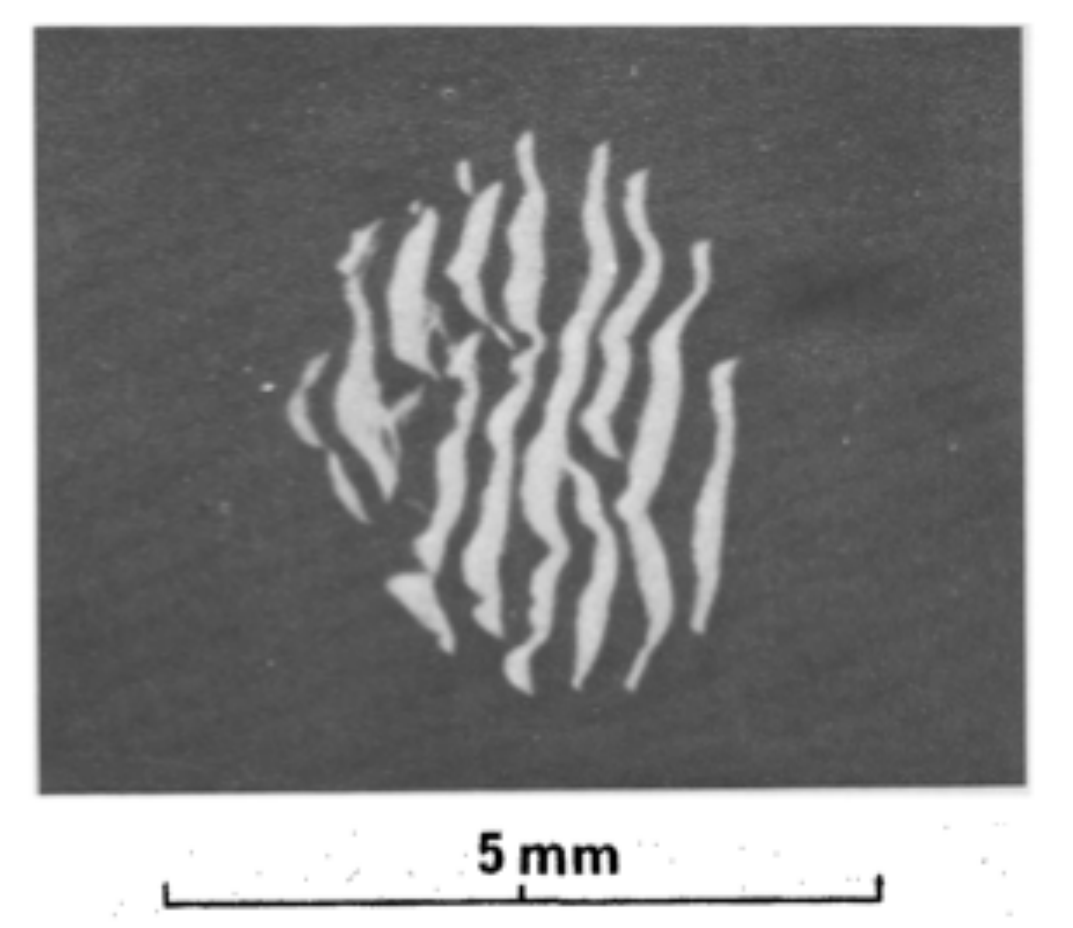}}
	\caption{{\small Straightening instabilities. (a) As the arm extends, wrinkles may appear on the surface of the elbow,  the number of which seems to depend mostly on the age of the subject. 
	(b) Single frame of a film of the contact between perspex and a butyl sphere sliding over it at 0.043 cm/s, showing the appearance of instability stripes (taken from \citet{Scha71}, permission yet to be sought).}}
\end{figure}

The back of the elbow on an extended arm and the region of contact between the road and a tyre are two examples of a \emph{straightening deformation}. In this paper we revisit and \color{black}  take \color{black} further the analysis of this somewhat neglected exact solution of nonlinear elasticity. 
We extend it to include a study of \emph{incremental instability}: in the case of the elbow, instability could model the appearance of wrinkles; in that of the tyre, it would correspond to the onset of the so-called  \citet{Scha71} stripes; see Figure \ref{fig0} for photographs.

\color{black}

There are only six known families of large deformations which are universal to all nonlinear incompressible isotropic materials (see for instance the  textbook \citep{TaME12} for a recent expos\'e). 
They are

\begin{itemize}
\item
\emph{Family 0}: Homogeneous deformations;
\item
\emph{Family 1}: Bending, stretching and shearing of a rectangular block;
\item
\emph{Family 2}: Straightening, stretching and shearing of a sector of a hollow cylinder;
\item
\emph{Family 3}: Inflation, bending, torsion, extension and shearing of an annular wedge; 
\item
\emph{Family 4}: Inflation or eversion of a sector of a spherical shell;
\item
\emph{Family 5}: Inflation, bending, extension and azimuthal shearing of an annular wedge.
\end{itemize}
There are countless stability studies for \emph{Families 0, 1, 3, 4}; however, there exists no work on the stability of a straightened sector. In fact, the literature on the large straightening deformation itself is very sparse and we have  been able to identify only five contributions on the subject \citep{Eric54, TrusNoll, Aron98, Aron00, Aron05}.

Although in practice, bending (\emph{Family 1}) and straightening (\emph{Family 2}) are the opposite of each other, their respective theoretical modelling differs completely. For instance, we shall see in this paper that the large straightening deformation occurs in plane strain and in plane stress ($\lambda_3=1$ and $\sigma_1=0$ throughout the block), while large bending is accompanied by plane strain only (the normal stress is zero only point-wise on the bent faces).
That no result at all can be deduced from one deformation to apply to the other is particularly true for the stability analysis, which proves extremely difficult to conduct.
Both problems can be formulated in terms of a linear ordinary differential system with variable coefficients which is stiff numerically. For the bending problem, this numerical stiffness can be smoothed out by the Compound Matrix method, which has proved most successful in the past in stability analyses for \emph{Family 0} (e.g. \cite{Haug11}), \emph{Family 1} (e.g. \cite{Haug99, CoDe08,  RoBG11}), \emph{Family 3} (e.g. \cite{DeOM10}) and \emph{Family 4} (e.g. \cite{Fu98, FuLi02}).
Here we tried several numerical methods in turn for the straightening problem: determinantal method, Compound Matrix method, Surface Impedance method, and only the latter one turned out to be robust enough to handle this very stiff problem (see also \citep{DeAC09, DOSV14b}).
\color{black}

\begin{figure}[!t]
\centering
\includegraphics[width=0.9\textwidth, keepaspectratio]{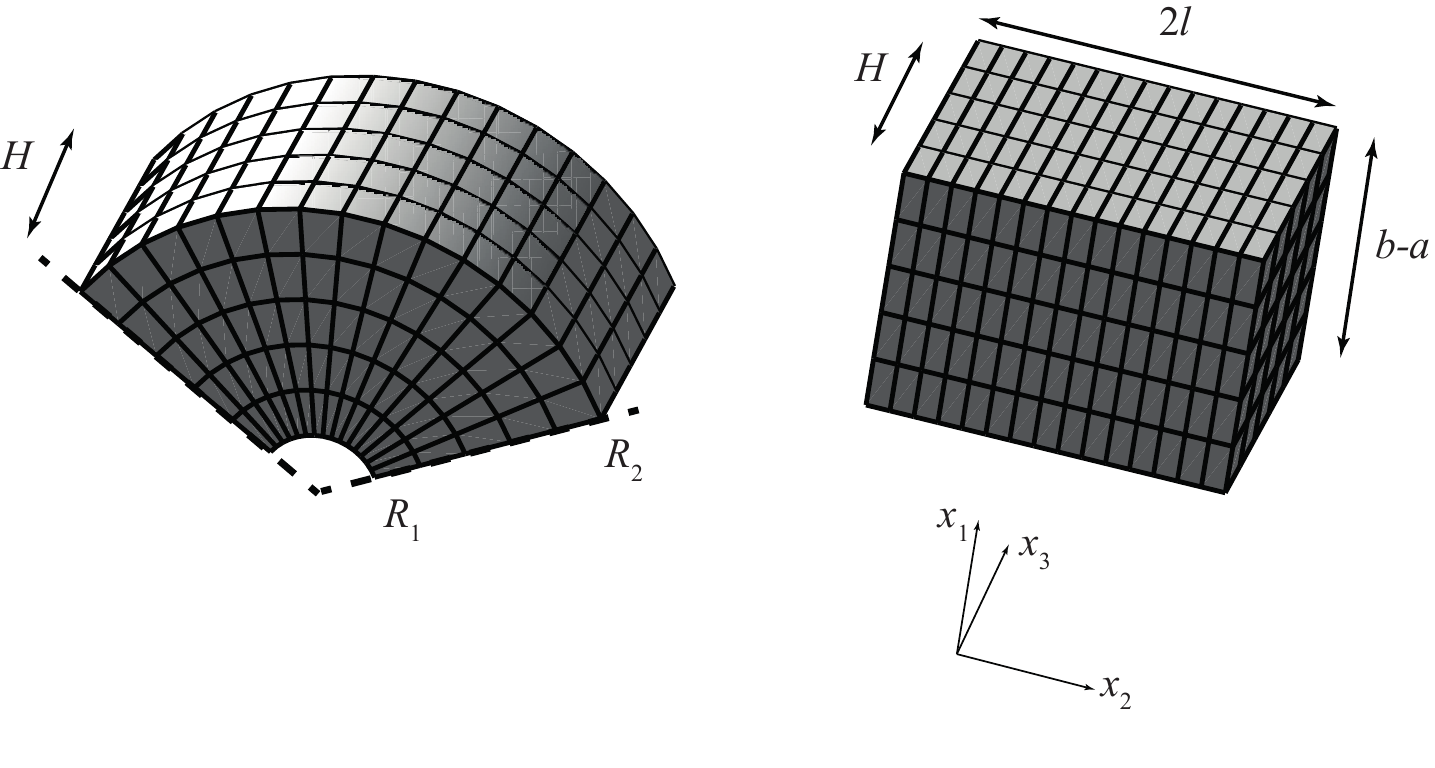}	
\caption{{\small Straightening of a circular cylindrical sector with initial radii ratio $R_1/R_2=0.2$ and open angle $4\pi/3$. 
The line elements along $x_2$ (originally circumferential) on the inner straightened face $x_1=a$ (originally at $R=R_1$) are extended; those on the outer face $x_1=b$ (originally at $R=R_2$) are contracted. Eventually, at a critical stretch of contraction, the outer face buckles and  wrinkles appear.}}\label{fig1}
\end{figure}

In his seminal paper on ``Deformations Possible in Every Isotropic, Incompressible, Perfectly Elastic Body'', J. L.  \citet{Eric54} showed that a circular sector can always be ``straightened'' into a rectangular block, by mapping the reference cylindrical  polar coordinates $(R, \Theta,Z)$ of the material points in the region 
\begin{equation}
0<R_1\leq R\leq R_2,\quad -\Theta_0 \leq \Theta\leq \Theta_0,\quad 0\leq Z\leq H,
\end{equation}
to the rectangular Cartesian coordinates $(x_1,x_2,x_3)$ of points in the region
\begin{equation}
a\leq x_1\leq b,\quad -l\leq x_2\leq l, \quad 0\leq x_3\leq H,
\end{equation}
in the deformed configuration through the deformation
\begin{equation} \label{deformation}
x_1=\frac{1}{2}AR^2,\quad x_2=\frac{\Theta}{A},\quad x_3=Z;
\end{equation}
see Figure \ref{fig1} and also  \citet{TrusNoll}.

Here $0 < 2\Theta_0 < 2\pi$ is the angle spanning the solid sector, and $2\pi-2\Theta_0$ is what we will call the ``open angle'' of the original, undeformed, sector (in contrast to the term ``opening angle'' of a deformed sector used, for example, in \citet{DeOM10}). 
Also, $R=R_1$ and $R=R_2$ are the inner and outer faces of the open sector, respectively, while $x_1=a$ and $x_1=b$ are their counterparts in the deformed rectangular block.  
The quantity $A$, which has the dimensions of the inverse of a length, is to be determined from the boundary conditions. 
In general, its value depends on the loads imposed on the end faces $x_2 = \pm l$ of the straightened block to sustain the deformation. 
Examples of such loads \color{black} (in the absence of body forces) \color{black} include a resultant normal force $N$, a resultant moment $M$, or a mixture of the two.  
In this paper we take the point of view that the final length $2l$ of the block is found from the stability analysis and we calculate the corresponding $A$, $N$ and $M$ required to attain it.  
In particular, it is clear from \eqref{deformation} that 
\begin{equation} \label{A}
A =\Theta_0/l,
\end{equation}
and that the faces normal to the $x_1$ axis are at 
\begin{equation} \label{faces}
x_1=a=\frac{1}{2}AR_1^2  =\frac{\Theta_0}{2l}R_1^2, \quad x_1=b=\frac{1}{2}AR_2^2  =\frac{\Theta_0}{2l}R_2^2.
\end{equation}

Section \ref{Basic equations} covers the characteristics of the large straightening deformation.
In particular, we show there that the principal Cauchy stress component $\sigma_1$ is zero throughout the block.

\begin{figure}[!h]
	\centering
		\includegraphics[width=0.85\textwidth]{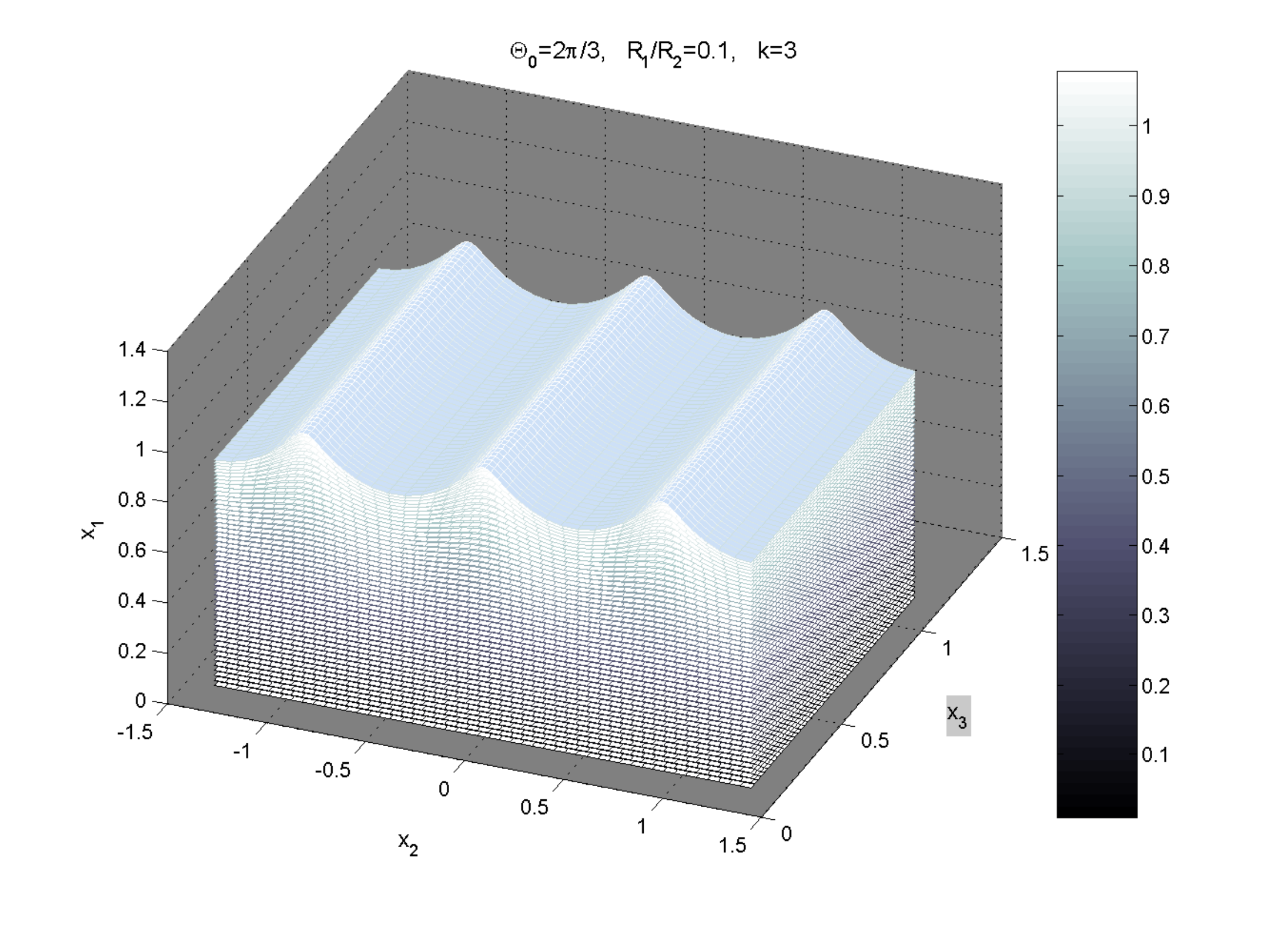}
	\caption{{\small When a sector made of Mooney--Rivlin material with angle $\Theta_0=2\pi/3$ and radii ratio $R_1/R_2=0.1$ is straightened, it buckles with the formation of $k=3$ wrinkles, with amplitude decaying rapidly away from the face $x_1=b$.}}
	\label{3D}
\end{figure}

Prescription of the final length $2l$ will prove to be most useful in the subsequent instability analysis (Section \ref{Stability analysis}), especially for making connections with known results of surface instability. 
Indeed, here the straightening instability threshold corresponds to the possible existence of static wrinkles on the face \eqref{faces}$_2$ of the straight block; see Figure \ref{3D} for a 3D representation of such wrinkles.

Consider the deformations on these surfaces: the face $R=R_2$ in the circumferential direction is deformed into the face $x_1=b$, and this face is compressed in the $x_2$ direction if $R_2\Theta_0>l$. This will be the case provided any axial tension applied to the faces $x_2=\pm l$ is not too large. Similarly, the face  $R=R_1$ will be stretched by the deformation if $R_1\Theta_0<l$.
This suggests, in particular, that on the face $x_1=b$ there should exist a \color{black} critical \color{black} compressive stretch, $\lambda_\text{cr}$ say, at which  wrinkles with  sinusoidal variations in the $x_2$-direction should appear, \color{black} and we shall use $\lambda_\text{cr}$ as our bifurcation parameter. \color{black}
The corresponding bifurcation criterion should be dispersive because of the presence of inherent characteristic lengths and, in particular, in the thick block/small wavelength limit it should yield the surface stability criterion of plane strain compression. 
For instance, for a block made of neo-Hookean, Mooney--Rivlin, or generic third-order incompressible elastic material, the asymptotic critical stretch should be $\lambda_\text{cr}=0.544$, as established by  \citet{Biot63}.

We conduct the full numerical analysis of straightening wrinkles in Section \ref{General stability analysis: numerics}, where we treat the example of the Mooney--Rivlin material (which here coincides with the neo-Hookean material since the deformation is restricted to plane strain).
In the preceding section (Section \ref{asymptotics}), we provide an asymptotic stability analysis for thin sectors of a general strain-energy density and establish the expression for the critical compressive stretch up to second order in the thickness \color{black} to radius ratio \color{black} of the sector.


\section{Large straightening deformation}
\label{Basic equations}


We compute the deformation gradient  $\mathbf{F}$ from \eqref{deformation} as
\begin{equation} \label{F}
\mathbf{F}=AR\mathbf{e}_1\otimes\mathbf{E}_R+\frac{1}{AR}\mathbf{e}_2\otimes\mathbf{E}_\Theta+\mathbf{e}_3\otimes\mathbf{E}_Z,
\end{equation}
where $\mathbf{E}_R,\mathbf{E}_\Theta,\mathbf{E}_Z$ and $\mathbf{e}_1,\mathbf{e}_2,\mathbf{e}_3$ are the cylindrical polar and Cartesian unit basis vectors in the reference and deformed  configurations, respectively.
It follows that the Eulerian principal directions of the deformation (defined as the directions of the eigenvectors of $\mathbf{F}\mathbf{F}^\mathrm{T}$) are the Cartesian basis vectors and that the
principal stretches (the square roots of the eigenvalues of $\mathbf{F}\mathbf{F}^\mathrm{T}$) are 
\begin{equation}
\lambda_1=AR,\quad \lambda_2=\frac{1}{AR},\quad \lambda_3=1,
\end{equation} 
showing that the deformation \eqref{deformation} is plane strain and isochoric.

In this paper, we consider homogeneous,  incompressible isotropic hyperelastic materials, for which the straightening deformation is universal  \citep{Eric54}. 
We denote by  $W=W(\lambda_1,\lambda_2,\lambda_3)$ the strain energy  per unit volume, so that the Cauchy stress tensor is 
\begin{equation}
\boldsymbol{\sigma}=\sigma_1\mathbf{e}_1\otimes\mathbf{e}_1+\sigma_2\mathbf{e}_2\otimes\mathbf{e}_2+\sigma_3\mathbf{e}_3\otimes\mathbf{e}_3,
\end{equation}
where $\sigma_1,\sigma_2,\sigma_3$ are the principal Cauchy stresses, given by
\begin{equation}\label{sigmas}
\sigma_1=\lambda_1\frac{\partial W}{\partial \lambda_1}-p,\quad \sigma_2=\lambda_2\frac{\partial W}{\partial \lambda_2}-p,\quad \sigma_3=\frac{\partial W}{\partial \lambda_3}-p,
\end{equation}
and $p$ is a Lagrange multiplier associated with the constraint of incompressibility ($\det \mathbf{F}=\lambda_1\lambda_2\lambda_3 \equiv 1$).

From now on it is convenient to introduce the notation 
\begin{equation}
\lambda_1 = \lambda^{-1}, \quad \lambda_2=\lambda,
\end{equation}
 and to define the strain energy in terms of a single deformation variable, the \emph{circumferential stretch} $\lambda$.  
Thus, we define the function $\hat W = \hat{W}(\lambda)$ by 
\begin{equation}
\hat{W}(\lambda)=W(\lambda^{-1},\lambda,1),
\end{equation}
and it then follows that 
\begin{equation}
\sigma_2-\sigma_1=\lambda\hat{W}'(\lambda),
\end{equation}
where the prime denotes the derivative with respect to $\lambda$.
We take the undeformed configuration to be stress free, so that 
\begin{equation}\label{undeformed}
\hat W'(1)=0.
\end{equation}

Clearly the deformation depends only on the single variable $R$ (or equivalently $x_1$) and hence the second and third components of the equilibrium equation $\div \boldsymbol{\sigma}=\mathbf{0}$ in the absence of body forces show that $p$ is independent of $x_2$ and $x_3$, while the first component yields simply $\mathrm{d}\sigma_1/\mathrm{d} x_1=0$. 
Hence the \emph{normal principal  stress component $\sigma_1$ is  constant throughout the block}.  

Next we assume that \color{black} the boundaries $x_1=a,b$ are free of traction, \color{black} and we deduce that 
\begin{equation} \label{sigma2}
\sigma_1\equiv 0,\quad \sigma_2=\lambda\hat{W}'(\lambda),
\end{equation}
while $\sigma_3$ has to be calculated from \eqref{sigmas}$_3$ with $p=\lambda_1\partial W/\partial \lambda_1$.  
We require the stress $\sigma_2$ to be positive when corresponding to stretching and negative when corresponding to contraction, and we therefore impose the conditions
\begin{equation}
\hat{W}'(\lambda)\gtreqqless 0\quad\mbox{according as}\quad \lambda \gtreqqless 1.\label{lambdaineqs}
\end{equation}
In view of \eqref{undeformed}, these conditions hold if $\hat{W}$ is a convex function of $\lambda$, and we assume this is the case henceforth.

Now we calculate the resultant normal force $N$ and moment $M$ (about the origin of the Cartesian coordinate system) on a  face $x_2=\mathrm{constant}$ of the block.  They are independent of $x_2$ and given by
\begin{equation}
N= H\int_a^b\sigma_2\mathrm{d}x_1,\quad M=- H\int_a^b\sigma_2x_1\mathrm{d}x_1.
\end{equation}
On use of the connections $\lambda=1/(AR)$ and $x_1=AR^2/2$, it is straightforward to show that
\begin{equation}\label{equation}
N = \dfrac{H}{A}\int_{\lambda_b}^{\lambda_a} \lambda^{-2} \hat{W}'(\lambda) \mathrm{d}\lambda,
\quad 
M = - \dfrac{H}{2A^2} \int_{\lambda_b}^{\lambda_a} \lambda^{-4} \hat{W}'(\lambda)\mathrm{d}\lambda,
\end{equation}
where
\begin{equation}
\lambda_a=\frac{1}{AR_1} >  \lambda_b=\frac{1}{AR_2},
\end{equation}
are the stretches in the $x_2$-direction on the faces $x_1=a$ and $x_1=b$ of the straightened block, respectively. 
In general, the resultant force  $N$ is not zero.
\color{black} Note that the possibility of straightening a circular cylindrical sector by applying a  moment alone and no normal force is treated in \citep{DOSV14b}.\color{black}


\section{Small  wrinkles}
\label{Stability analysis}


In order to study the (linearized)  stability of the deformed rectangular configuration, we investigate the possible existence of infinitesimal static solutions in the neighbourhood of the large straightening deformation. Hence we consider superimposed displacements $\mathbf{u} = \mathbf u(x_1, x_2, x_3)$.  
We let $\mathbf{L} = \grad\mathbf{u}$ denote the displacement gradient, so that the incremental incompressibility condition reads
\begin{equation}\label{incremental incompressibility}
\tr\mathbf{L}\equiv L_{ii}=u_{i,i}=0
\end{equation}
in the usual summation convention for repeated indices, where a subscript $i$ following a comma signifies differentiation with respect to $x_i$.

The incremental nominal stress  $\mathbf{\dot{s}_0}$ is given by \citep{Ogde97}
\begin{equation}\label{incremental stress}
\mathbf{\dot{s}_0} = \boldsymbol{\mathcal{A}_0} \mathbf{L}+p\mathbf{L}-\dot{p}\mathbf{I},
\end{equation}
where a superposed dot signifies an increment, the zero subscript indicates evaluation in the deformed configuration, $\mathbf{I}$ is the identity tensor, and $\boldsymbol{\mathcal{A}_0}$ is the fourth-order tensor of instantaneous elastic moduli.  In components
\begin{equation} \label{S0}
\dot{s}_{0ij}=\mathcal{A}_{0ijkl}u_{l,k}+pu_{i,j}-\dot{p}\delta_{ij},
\end{equation}
where $\delta_{ij}$ is the Kronecker delta.  Referred to the (Eulerian) principal axes of the underlying deformation, the only non-zero components of $\boldsymbol{\mathcal{A}_0}$ are
\begin{align}
& \mathcal{A}_{0iijj}=\lambda_i\lambda_jW_{ij}, \quad i,j = 1,2,3, \notag
\\[3pt]
& \mathcal{A}_{0ijij}=\mathcal{A}_{0ijji}+\lambda_iW_i= (\lambda_iW_i-\lambda_jW_j)\dfrac{\lambda_i^2}{\lambda_i^2-\lambda_j^2}, \quad i\neq j,\quad \lambda_i\neq\lambda_j,\label{moduli}
\end{align}
where $W_i \equiv \partial W/\partial\lambda_i$, $W_{ij} \equiv \partial^2W/\partial\lambda_i\partial\lambda_j$ (note the major symmetry $\mathcal{A}_{0piqj}=\mathcal{A}_{0qjpi}$).  The  corresponding formulas for $\lambda_j=\lambda_i,\,i\neq j$, in \eqref{moduli}$_2$ may be obtained by a limiting procedure,  but are not needed here.

In the absence of body forces, the incremental equation of equilibrium has the form 
\begin{equation}\label{eq}
\div\: \mathbf{\dot{s}_0} = \mathbf{0}.
\end{equation}
Here, for  simplicity, we focus on plane two-dimensional incremental deformations such that $u_3=0$ and $u_1$, $u_2$ are independent of $x_3$. Further, we seek solutions which have sinusoidal variations in the $x_2$-direction, i.e. of the form
\begin{equation}\label{assumption}
\{u_1,u_2,\dot{p}\}=\{U_1(x_1),U_2(x_1),P(x_1)\}\mathrm{e}^{\im nx_2},
\end{equation}
\color{black}
where 
\begin{equation} \label{n}
n = k\pi A/\Theta_0 = k\pi/l,
\end{equation}
is the wavenumber,  and the integer  $k$ is the \emph{mode number}, identifying the number of wrinkles on the faces $x_1 =$ constant, with amplitude decaying from $x_1=b$ to $x_1=a$.
Clearly, the \emph{wavelength} $\mathscr L$ of the wrinkles is
\begin{equation}\label{L}
\mathscr L = 2\pi/n = 2l/k,
\end{equation}
i.e. it is equal to the length of the deformed block divided by the number of wrinkles. 
These two quantities are to be determined from the numerical treatment of the stability analysis.
\color{black}

It follows from \eqref{S0} that the components of the incremental nominal stress tensor have a similar form, \begin{equation}
\dot{s}_{0ij}=S_{ij}(x_1)\mathrm{e}^{\im nx_2}, \quad i,j=1,2,
\end{equation}
say.
Then, by using a standard procedure (see, e.g.,  \citet{DeSc04,DeOg05,DeAC09, DeOM10}), we  cast the above equations as a first-order differential system for the  four-component displacement--traction vector  $\boldsymbol{\eta} \equiv [U_1,U_2,\im  S_{11},\im S_{12}]^\text{T}$.
This is the so-called \emph{Stroh formulation}:
\begin{equation}\label{stroh}
\frac{\mathrm{d}}{\mathrm{d} x_1}\boldsymbol{\eta}(x_1)=\im \: \mathbf{G}(x_1)\boldsymbol{\eta}(x_1)
=\im \: \begin{pmatrix} \mathbf{G_1}(x_1) & \mathbf{G_2}(x_1) \\ \mathbf{G_3}(x_1) & \mathbf{G_1}(x_1) \end{pmatrix}\boldsymbol{\eta}(x_1).
\end{equation}
Here the real matrix $\mathbf{G}$ has the form
\begin{equation}\label{gi}
\mathbf{G}=\left(\begin{array}{cccc}
0 & -n & 0 & 0\\
[3mm]
-n & 0 & 0 & -1/\alpha\\
[3mm]
n^2\sigma_2 & 0 & 0 & -n\\
[3mm]
0 & n^2\lambda^2\hat{W}''(\lambda) & -n & 0
\end{array}\right),
\end{equation}
where
\begin{equation}
 \alpha=\mathcal{A}_{01212}= \lambda\hat{W}'(\lambda)/(\lambda^4-1),\label{alpha-definition}
\end{equation}
and its $2 \times 2$ sub-matrices $\mathbf{G_1},\mathbf{G_2},\mathbf{G_3}$ are symmetric.

Now, if the incremental equations of equilibrium can be solved, subject to the traction-free conditions 
\begin{equation}\label{bcxi}
S_{11}=S_{12}=0 \quad \mathrm{on}\quad x_1=a,b,
\end{equation}
then possible equilibrium states exist in a neighbourhood of the straightened configuration, signalling the \emph{onset of instability}. 
When such a configuration is determined, we refer to the value of the circumferential stretch $\lambda_b=1/(AR_2)$ at this point as the \emph{critical} value for compression, which we denote by  $\lambda_\text{cr}$.

Simple calculations show that
\begin{equation}
A(b-a)=\frac{R_2^2-R_1^2}{2\lambda_\text{cr}^2R_2^2},
\end{equation}
which allows for the complete determination of the straightened geometry just prior to instability. In particular, the dimensions of the rectangular block in the $x_1$ and $x_2$ directions are then, respectively, 
\begin{equation}\label{dimensions}
b-a=\frac{R_2^2-R_1^2}{2\lambda_\text{cr}R_2},\quad \color{black} l=R_2 \Theta_0 \lambda_\text{cr}.\color{black}
\end{equation}
\color{black}
This latter equation, coupled to \eqref{L}, shows that the critical wavelength of bifurcation is the \color{black} length of the external curved boundary of the undeformed \color{black} sector, multiplied by the critical stretch ratio and divided by the number of wrinkles.
\color{black}

Take, for instance, the sector in Figure \ref{fig1}(a). 
To draw it we scaled lengths with respect to $R_2$ and picked $R_1=0.2$,  $R_2=1$, $\Theta_0=\pi/3$. 
The stability analysis of Section \ref{General stability analysis: numerics} will reveal that if the material is modelled by the Mooney--Rivlin potential, then the corresponding buckled state occurs at $\lambda_\text{cr} = 0.5710$, \color{black} with $k=1$ wrinkle forming over the surface\color{black}; see Figure \ref{last-fig}.
It follows from \eqref{dimensions} that just prior to instability, at $\lambda_b=0.5711$ say, the dimensions of the rectangular block are $b-a = 0.840$ and $l= 0.598$; see  Figure \ref{fig1}(b).
Then $A$ can be found from $\lambda_\text{cr}$, and $N$ and $M$ from \eqref{equation}.
\color{black} By \eqref{L}, the critical wavelength is $\mathscr L = 1.20$.\color{black}


\section{Asymptotic analysis for thin sectors}
\label{asymptotics}


In this section we derive the approximation to the critical threshold $\lambda_\text{cr}$ when the undeformed cylindrical sector is  thin compared to the radius of the outer face, that is when
\begin{equation}
\eps \equiv 1 - \frac{R_1}{R_2}\ll1.
\end{equation}

To conduct a perturbation analysis of the incremental equations of equilibrium \eqref{stroh}--\eqref{bcxi} in terms of the small parameter $\eps$, we need to expand the strain energy in terms of the strain. 
All expansions are equivalent, and here we choose to use the Green--Lagrange strain tensor $\mathbf E = (\mathbf F^\text{T}\mathbf F - \mathbf I)/2$.
For incompressible isotropic elastic solids, the expansion begins as  
\begin{equation}
W=\mu\,  \tr(\mathbf{E}^2) + \dfrac{\mathcal A}{3}\,  \tr(\mathbf{E}^3) +\mathcal D \left(\tr(\mathbf{E}^2)\right)^2 + \ldots,
\end{equation}
where $\mu$, $\mathcal A$, $\mathcal D$, the respective second-, third-, and fourth-order elastic constants, are all of the same order of magnitude and of the same dimensions (see \citet{DeOg10} and references therein).
For the straightening deformation \eqref{F}, indeed for any plane strain deformation, we obtain for future reference the results
\begin{align}\label{reduction}
& \hat{W}'(1) = 0, \quad
\hat{W}''(1)= 4\mu, \quad
\hat{W}'''(1) = -12\mu, \notag \\ 
& \hat{W}''''(1) = 156 \mu + 48 \mathcal{A} + 96\mathcal{D}, \quad  
\alpha(1)=\mu, \quad 
\alpha'(1) = -2\mu,
\end{align}
\color{black} where $\alpha=\alpha(\lambda)$ is defined by \eqref{alpha-definition}.\color{black}

We now introduce the following \emph{dimensionless quantities}
\begin{align}\label{dimless}
& \displaystyle \xi=\frac{1}{\eps}\left(1-\sqrt{\frac{2x_1\lambda_\text{cr}}{R_2}}\right), \quad 
\bar n=\frac{k\pi}{\Theta_0}, \quad 
\bar \alpha = \dfrac{\alpha}{\mu},\notag \\[3pt]
& \displaystyle \bar \sigma_2 = \dfrac{\sigma_2}{\mu},\quad 
\bar U_i = \dfrac{U_i}{R_2}, \quad 
\bar S_{1i} = \dfrac{S_{1i}}{\mu}, \quad
\bar W = \dfrac{\hat W}{\mu}.
\end{align}
The space variable is thus $\xi$, which has the advantage of spanning the thickness of the block with the fixed range $[0,1]$.
In terms of the new variable $\xi$, the stretch $\lambda$ reads as
\begin{equation}\label{lambda}
\lambda=\frac{\lambda_\text{cr}}{1-\eps \xi}.
\end{equation}
Substituting (\ref{dimless}) into \color{black} Equations \color{black} (\ref{stroh})--(\ref{gi}) yields
\begin{equation}\label{dimless equation}
\frac{\mathrm{d}\boldsymbol{\bar \eta}}{\mathrm{d}\xi}=-\im\eps\lambda^{-1}\mathbf{\bar G}\boldsymbol{\bar \eta},
\end{equation}
where $\boldsymbol{\bar \eta} =[\bar U_1, \bar U_2,\im \bar S_{11}, \im  \bar S_{12}]^\text{T}$ and the dimensionless Stroh matrix is
\begin{equation}\label{Gstar}
\mathbf{\bar G}=\left(\begin{array}{cccc}
0 & - \bar n /\lambda_\text{cr} & 0 & 0\\
[3mm]
- \bar n/\lambda_\text{cr} & 0 & 0 & -1/\bar \alpha\\
[3mm]
\bar n^{2}\bar \sigma_2 / \lambda_\text{cr}^{2}& 0 & 0 & - \bar n/ \lambda_\text{cr}\\
[3mm]
0 & \bar n^{2} \bar W''/(1-\eps\xi)^2 & - \bar n/\lambda_\text{cr} & 0
\end{array}\right).
\end{equation}
Finally, we introduce the power series expansions of $\lambda_\text{cr}$ and of  $\boldsymbol{\eta}$ in $\eps$:
\begin{equation}\label{bc}
\lambda_\text{cr} = \sum_{i=0}^{\infty} \eps^i \lambda^{(i)}_\text{cr}, \quad 
\boldsymbol{\bar \eta}=\sum_{i=0}^{\infty}\eps^i\boldsymbol{\eta^{(i)}}.
\end{equation}
The dimensionless displacement--traction vector  $\boldsymbol{\bar \eta}$ must satisfy the boundary conditions \eqref{bcxi} at all orders, so that \begin{equation}\label{bch}
\eta_{3}^{(i)} = \eta_{4}^{(i)}=0, \quad \textrm{for}\quad \xi=0,1.
\end{equation}

We start the asymptotic analysis at order $\mathcal O(1)$, where Equation \eqref{dimless equation} gives
\begin{equation}
\frac{\d\boldsymbol{\eta^{(0)}}}{\d \xi}=\mathbf{0},
\end{equation}
for which the solutions satisfying \eqref{bch} are of the form
\begin{equation}\label{order0}
\boldsymbol{\eta^{(0)}}=[a_0, b_0, 0, 0]^\text{T},
\end{equation}
where $a_0$ and $b_0$ are integration constants, to be determined at the next order of the asymptotic analysis.

At order $\mathcal O(\eps)$, Equation \eqref{dimless equation} gives 
\begin{equation}\label{order1}
\dfrac{\d\boldsymbol{\eta^{(1)}}}{\d \xi}=-\im (\lambda_\text{cr}^{(0)})^{-1}\mathbf{G^{(0)}}\boldsymbol{\eta^{(0)}},
\end{equation}
with
\begin{equation}\label{Gstar0}
\mathbf{G^{(0)}}=\left(\begin{array}{cccc}
0 & - \bar n/\lambda_\text{cr}^{(0)} & 0 & 0\\
[3mm]
- \bar n/ \lambda_\text{cr}^{(0)} & 0 & 0 & - 1/\bar \alpha\\
[3mm]
\bar n^{2}\bar{W}' /\lambda_\text{cr}^{(0)}& 0 & 0 & - \bar n/\lambda_\text{cr}^{(0)}\\
[3mm]
0 & \bar n^{2}\bar{W}'' & - \bar n/\lambda_\text{cr}^{(0)}& 0
\end{array}\right),
\end{equation}
where $\bar \alpha$, $\bar W'$ and $\bar W''$ are evaluated at $\lambda_\text{cr}^{(0)}$
(note that we have used the connection \eqref{sigma2}$_2$ to express $\bar \sigma_2$ in terms of $\bar W'$).
By direct integration of \eqref{order1} we obtain
\begin{align}\label{solution}
& \eta_{1}^{(1)} = -\im (\lambda_\text{cr}^{(0)})^{-1}\left[ a_1- (\bar n /\lambda_\text{cr}^{(0)})b_0 \xi\right],\notag \\
&\eta_{2}^{(1)} = -\im (\lambda_\text{cr}^{(0)})^{-1}\left[ b_1- (\bar n /\lambda_\text{cr}^{(0)})a_0 \xi\right],\notag \\
&\eta_{3}^{(1)} = -\im (\lambda_\text{cr}^{(0)})^{-1}(\bar n^{2}\bar{W}' /\lambda_\text{cr}^{(0)})a_0\xi,\notag \\
&\eta_{4}^{(1)} = -\im ( \bar n^{2}\bar{W}'' )b_0\xi,
\end{align}
where $a_1$ and $b_1$ are constants of integration for $\eta_{1}^{(1)}$ and $\eta_{2}^{(1)}$, respectively, while the constants of integration for $\eta_{3}^{(1)}$ and $\eta_{4}^{(1)}$ were taken as zero in order to satisfy \eqref{bch} at $\xi=0$.  
To satisfy the boundary condition at $\xi=1$ we must have, simultaneously, \{\emph{either} $\bar W' (\lambda_\text{cr}^{(0)})=0$ \emph{or} $a_0=0$\} and \{\emph{either} $\bar W'' (\lambda_\text{cr}^{(0)})=0$ \emph{or} $b_0=0$\}.
As a  consequence of \eqref{lambdaineqs} and of the convexity of $\bar W$, the unique  value of $\lambda_\text{cr}^{(0)}$ satisfying this is $\lambda_\text{cr}^{(0)} = 1$, which is in line with our expectation that sectors of vanishing thickness should buckle as soon as deformed (see \citet{CoDe08} or \citet{GoDV08} for similar conclusions for the bending or compression of thin elements). 
Using  \eqref{reduction}, we thus obtain the following reduced expressions for the displacement--traction vector:
\begin{equation}
\boldsymbol{\eta^{(0)}}=[a_0,0,0,0]^\text{T}, \quad \boldsymbol{\eta^{(1)}} =-\im[a_1,b_1- \bar n a_0\xi,0,0]^\text{T};
\end{equation}
and for the matrix $\mathbf{G^{(0)}}$:
\begin{equation}\label{Gstar0new}
\mathbf{G^{(0)}} = 
\left(\begin{array}{cccc}
0 & - \bar n & 0 & 0\\
[3mm]
- \bar n & 0 & 0 & -1\\
[3mm]
0 & 0 & 0 & - \bar n\\
[3mm]
0 & 4 \bar n^{2} & - \bar n & 0
\end{array}\right).
\end{equation}

Moving on to the collection of the terms of order $\mathcal O(\eps^2)$ in Equation \eqref{dimless equation}, we obtain
\begin{equation}\label{order2}
\dfrac{\d\boldsymbol{\eta^{(2)}}}{\d\xi}
 = -\im \left[ \mathbf{G^{(1)}}\boldsymbol{\eta^{(0)}} + \mathbf{G^{(0)}} \boldsymbol{\eta^{(1)}} - (\lambda_\text{cr}^{(1)}+\xi)\mathbf{G^{(0)}}\boldsymbol{\eta^{(0)}} \right],
\end{equation}
where  the matrix $\mathbf{G^{(0)}}$ is given by \eqref{Gstar0new} from now on, and 
\begin{equation}
\mathbf{G^{(1)}}=\left(\begin{array}{cccc}
0 & \bar n \lambda_\text{cr}^{(1)} & 0 & 0\\
[3mm]
\bar n \lambda_\text{cr}^{(1)} & 0 & 0 & -2( \lambda_\text{cr}^{(1)} +\xi)\\
[3mm]
4 \bar n^2 (\lambda_\text{cr}^{(1)} + \xi) & 0 & 0 &\bar n \lambda_\text{cr}^{(1)} \\
[3mm]
0 & -4 \bar n^2 \left(\xi + 3\lambda_\text{cr}^{(1)}  \right) &\bar n \lambda_\text{cr}^{(1)}  & 0
\end{array}\right),
\end{equation}
where again, we have used  \eqref{reduction}.
We  then integrate these equations directly and apply the boundary conditions to deduce $\lambda_\text{cr}^{(1)}$, and repeat the process at the next order. 
The results are obtained simply as
\begin{equation}\label{first}
\lambda_\text{cr}^{(1)} = -\frac{1}{2}, \quad 
\lambda_\text{cr}^{(2)} = -\frac{1}{24}\left( 2 \bar n^2-3\right).
\end{equation}

Here we see that the correction at order one to the critical stretch is the same for any cylindrical sector irrespective of its open angle, and is valid for any value of the mode number $k$. 
The effects of the open angle and of the mode number on the critical threshold $\lambda_\text{cr}$ reveal themselves at order two in the thickness, although that correction accounts only for geometrical effects since no physical elastic constant is present. 
Clearly, the largest value of $\lambda_\text{cr}^{(2)}$ corresponds to $\bar n=\pi/\Theta_0$, that is when $k=1$ 
and only one vertical wrinkle occurs on the outer face of the straightened block.

We have thus established the following \emph{universal} asymptotic expansion of the critical compression value of the circumferential stretch,
 \begin{equation}\label{lambdacrnew}
\lambda_\text{cr}=1- \frac{1}{2}\eps - \frac{1}{24}\left(2\dfrac{\pi^2}{\Theta_0^2}-3\right)\eps^2+ \mathcal O(\eps^3).
\end{equation}


\section{General stability analysis: numerics}
\label{General stability analysis: numerics}


The inhomogeneous differential system \eqref{stroh} is numerically stiff and calls for a robust algorithm.
Here, we favour the \emph{Impedance Matrix method} (see  \citet{DeAC09} and \citet{norris12} for successful implementation of this algorithm in solids with cylindrical symmetry, and see  \citet{Shuv03} for a rigorous and exhaustive treatment of the underlying theory).

To summarize, the conditional impedance matrix $\mathbf{Z_a}$ is zero on the ``inner'' face $x_1=a$ and singular on the ``outer'' face $x_1=b$:
\begin{equation} \label{BCz}
\mathbf{Z_a}(a) = \mathbf 0, \quad
\det \mathbf{Z_a}(b) = 0.
\end{equation}
By definition it relates the displacement $\mathbf U$ to the traction $\mathbf S$ through 
\begin{equation}\label{derivation}
\mathbf{S}(x_1)=\mathbf{Z_a}(x_1)\mathbf{U}(x_1),
\end{equation}
and by construction it satisfies the matrix differential Riccati equation
\begin{equation}\label{riccati}
\frac{\mathrm{d}\mathbf{Z_a}}{\mathrm{d}x_1} = 
 \im(\mathbf{G_1}\mathbf{Z_a} - \mathbf{Z_a} \mathbf{G_1}) + \mathbf{Z_a} \mathbf{G_2} \mathbf{Z_a} + \mathbf{G_3}, 
\end{equation}
where the $\mathbf G$'s are  $2\times2$ sub-blocks of the Stroh matrix \eqref{gi}.

To solve numerically the boundary value problem given by \eqref{riccati} 
with boundary conditions \eqref{BCz} and to identify the critical value 
$\lambda_\text{cr}$, we apply a \emph{shooting}-like technique that 
combines the bisection (dichotomy) method with an initial value solver 
for \eqref{riccati} with \eqref{BCz}$_1$.
Hence, for a given value of $\lambda_\text{cr}$ in some starting guess 
interval,  we integrate \eqref{riccati} subject to \eqref{BCz}$_1$ by 
using the \texttt{ode45} function of Matlab (based on Runge--Kutta 
methods) and, if the target value \eqref{BCz}$_2$ is not met at $x_1=b$, 
then by dichotomy programming we adjust $\lambda_\text{cr}$ until the 
target is satisfied with a given precision ($\mathrm{tol}=10^{-7}$).

Then, once the critical value of the circumferential stretch is found, we have $\mathbf S(b)=\mathbf{Z_a}(b)\mathbf{U}(b)=\mathbf{0}$, which  means that
\begin{equation} \label{shape}
\frac{U_2(b)}{U_1(b)}=-\frac{Z_{a11}(b)}{Z_{a12}(b)}=-\frac{Z_{a21}(b)}{Z_{a22}(b)}.
\end{equation}
This ratio determines the shape of the \color{black}wrinkles \color{black} on the outer face of the straightened block.
Moreover, the displacement $\mathbf U$ satisfies the differential equation
\begin{equation} \label{U}
\dfrac{\mathrm{d} \mathbf{U}}{\mathrm{d}x_1}= \im\mathbf{G_1}\mathbf{U} - \mathbf{G_2}\mathbf{Z_a}\mathbf{U},
\end{equation}
and we may thus use \eqref{shape} as the starting point for its backward (from face $x_1=b$ to face $x_1=a$) numerical integration in order to compute the entire displacement field (a similar equation also exists for the traction field).

For our numerical experiments, we first  non-dimensionalize the equations (note that this is a different non-dimensionalization process from that conducted in the asymptotic analysis of the previous section). 
Specifically, we use the dimensionless quantities
\begin{align}\label{dimlessx}
& x_1^* = \dfrac{x_1}{b}\in[R_1^2/R_2^2,1], \quad n^*=\dfrac{k\pi}{2\Theta_0}, \quad U_i^*=\frac{U_i}{b},   \quad \mathbf{Z^*_a} = \dfrac{b}{\mu}\mathbf{Z_a},  \notag \\[5mm]
&  \alpha^*=\frac{\alpha}{\mu}, \quad W^*=\dfrac{\hat W}{\mu}, \quad \displaystyle \sigma_2^*=\frac{\sigma_2}{\mu},\quad   S_{1j}^*=\frac{S_{1j}}{\mu}, 
\end{align}
where $\mu = \hat W''(1)/4$ is the infinitesimal shear modulus. 
Note that by using \eqref{deformation}, \eqref{A} and  $\lambda_\text{cr}$, it becomes a simple matter to establish the useful links
\begin{equation}
x_1^* = \lambda^{-2}\lambda_\text{cr}^2, \quad
n^*=b\lambda_\text{cr}^2n.
\end{equation}
Then, substitution of \eqref{dimlessx} into Equation \eqref{stroh} yields a starred version of the governing equations \eqref{stroh}, \eqref{riccati}, \eqref{shape} and \eqref{U}, where the non-dimensional\-ized Stroh matrix $\mathbf{G^*}$ is made up of the following blocks
\begin{align}
& \mathbf{G_1^*} = b\mathbf{G_1}=\left(\begin{array}{cc}
0 & -n^*\lambda_\text{cr}^{-2}\\
-n^*\lambda_\text{cr}^{-2} & 0
\end{array}\right), \quad
\mathbf{G_2^*} = \mu \mathbf{G_2} =
\left(\begin{array}{cc}
0 & 0\\
0 & -1/\alpha^*
\end{array}\right), \notag \\[12pt]
& \mathbf{G_3^*} =  \dfrac{b^2}{\mu} \mathbf{G_3} = \left(\begin{array}{cc}
n^{*2}\lambda_\text{cr}^{-4}\sigma_2^* & 0\\
0 & n^{*2} \lambda_\text{cr}^{-4} \lambda^2W^{*''}
\end{array}\right),
\end{align}
and the boundary conditions \eqref{BCz} are replaced by
\begin{equation}\label{dimless riccati}
 \mathbf{Z^*_a}(R_1^2/R_2^2)=\mathbf{0}, \quad \det\mathbf{Z_a^*}(1)=0.
\end{equation}

Next, we focus on a specific material for the constitutive modelling, namely the Mooney--Rivlin material, with strain energy density
\begin{equation}
W = C_1[\text{tr}(\mathbf{FF}^\text{T}) -3] + C_2[\text{tr}(\mathbf{FF}^\text{T})^{-1}-3],
\end{equation}
where $C_1$, $C_2$ are positive constants, \color{black} with $C_1+C_2=2\mu$\color{black}.
Then we find that
\begin{equation}
\alpha^*= \lambda_\text{cr}^{-2} x_1^*, 
\quad 
\sigma_2^*= \dfrac{\lambda_\text{cr}^2}{x_1^*} - \dfrac{x_1^*}{\lambda_\text{cr}^2},
\quad 
\lambda^2W^{*''} =  \dfrac{\lambda_\text{cr}^2}{x_1^*} +3 \dfrac{x_1^*}{\lambda_\text{cr}^2},
\end{equation}
and the non-dimensional equations can therefore be solved independently of the physical constants $C_1$ and $C_2$ (see \citet{DeAC09} for a study of the influence of constitutive parameters on the stability of bent blocks).

The quantities left at our disposal are geometric: the angle $\Theta_0$ and $k$, the number of wrinkles on the face $x_1=b$ (i.e. $x_1^*=1$).
Each choice of these quantities determines $n^*$ and then, for each initial thickness ratio $R_1/R_2$, we seek the corresponding critical stretch $\lambda_\text{cr}$. 

\begin{figure}[!h]
	\centering
		\includegraphics[width=0.95\textwidth]{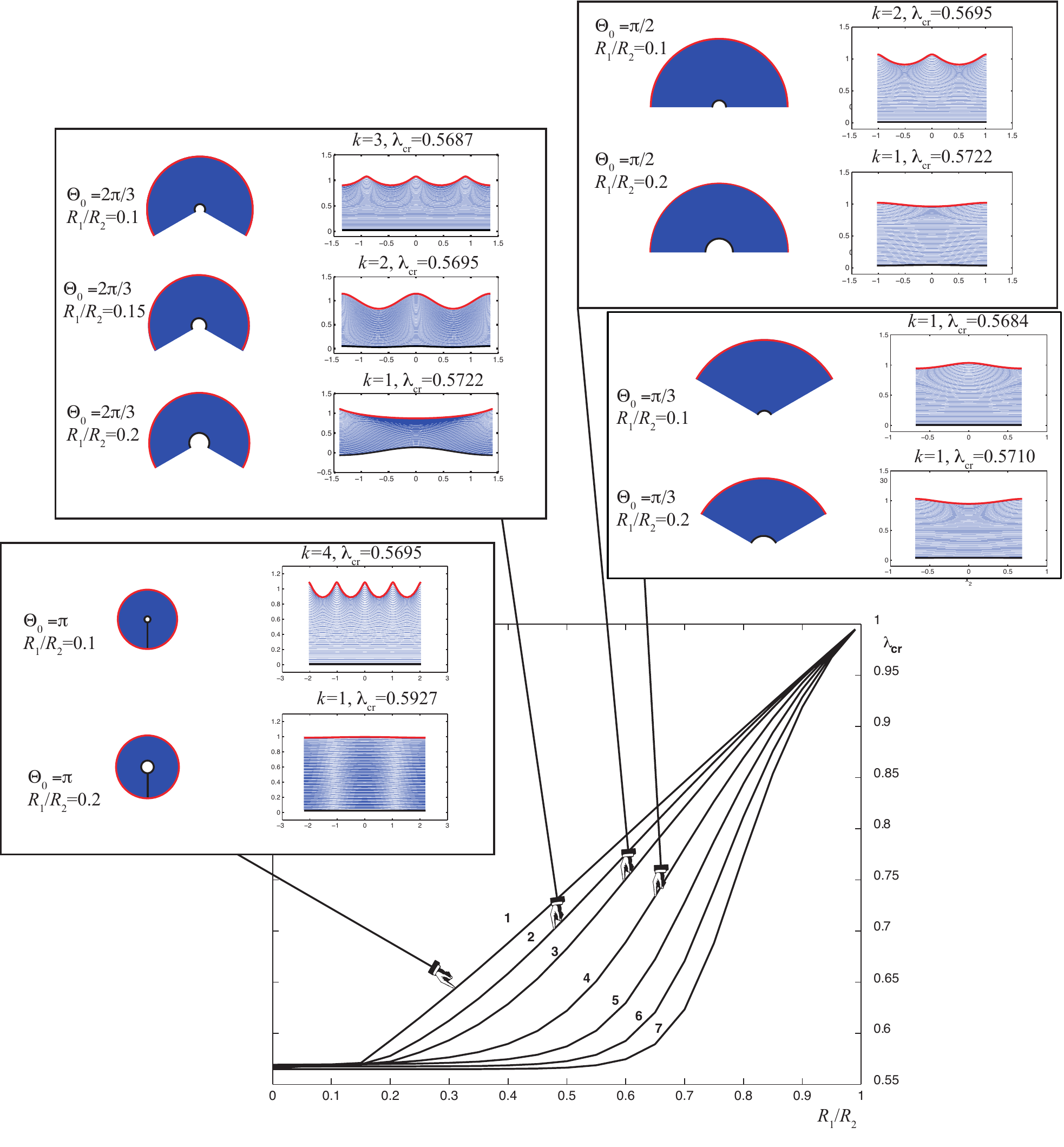}
	\caption{{\small Some examples of \color{black} the \color{black} critical value of the stretch $\lambda_\text{cr}$ as a function of the initial radii ratio $R_1/R_2$ for a Mooney--Rivlin material. The number of wrinkles $k$ depends on the  angle $\Theta_0$ and \color{black} on  $R_1/R_2$; \color{black} see Table 1 for details. For instance, for $\Theta_0=2\pi/3$ (curve 2) the number of wrinkles is $k=3$, 2, or 1, depending on the value of $R_1/R_2$. For $\Theta_0= \pi/3, \pi/4, \pi/5, \pi/6$ (curves $4,5,6,7$)  there is only one wrinkle, for any radii ratio.}}
\label{last-fig}
	\end{figure}

For practical purposes here, we choose a given angle $\Theta_0$ and vary $k$ to ensure that the value of $\lambda_\text{cr}$ corresponds to the earliest onset of buckling (i.e. $\lambda_\text{cr}$ is as close to 1 as possible).
Take for instance a sector with angle $\Theta_0=\pi/2$. 
We find that for $0 \le R_1/R_2 \le 0.1833$, the straightened block buckles with $k=2$ wrinkles, and for $0.1833 \le R_1/R_2 \le 1$, it buckles with $k=1$ wrinkle.
 \color{black}  Several different examples are listed in Table 1 and displayed in \color{black} Figure \ref{last-fig}.
In the thick sector (i.e. $R_1/R_2$ close to 0), small wavelength (i.e. $\Theta_0$ close to 0) limit, all curves tend to $\lambda_\text{cr} = 0.544$, the value found by \citet{Biot63} for surface instability in plane strain of a Mooney--Rivlin half-space (note that for plane strain the Mooney--Rivlin energy function reduces to the neo-Hookean one). 
In the thin sector limit, all curves are approximated by the quadratic \eqref{lambdacrnew}.

\begin{table}[!t]
\begin{center}
    \begin{tabular}{|c|c|l|}
        \hline
        Curve &  & ~ \\
        number & Angle & Number of wrinkles \\ \hline
        1 & $\Theta_0=\pi$ & $k=4$ for $0 \le R_1/R_2 \le 0.15$ \\ 
        ~ & ~ & $k=1$ for $0.15 \le R_1/R_2 \le 1$ \\ \hline
     2 & $\Theta_0=2\pi/3$ & $k=3$ for $0 \le R_1/R_2 \le 0.11$ \\ 
        ~ & ~ & $k=2$ for $0.11 \le R_1/R_2 \le 0.17$ \\
          ~ & ~ & $k=1$ for $0.17 \le R_1/R_2 \le 1$ \\ \hline
      3 & $\Theta_0=\pi/2$ & $k=2$ for $0 \le R_1/R_2 \le 0.18$ \\ 
        ~ & ~ & $k=1$ for $0.18 \le R_1/R_2 \le 1$ \\ \hline
      4 & $\Theta_0=\pi/3$ & $k=1$ for $0 \le R_1/R_2 \le 1$ \\ \hline
      5 & $\Theta_0=\pi/4$ & $k=1$ for $0 \le R_1/R_2 \le 1$ \\ \hline
       6 & $\Theta_0=\pi/5$ & $k=1$ for $0 \le R_1/R_2 \le 1$ \\ \hline
           7 & $\Theta_0=\pi/6$ & $k=1$ for $0 \le R_1/R_2 \le 1$ \\ \hline
    \end{tabular}
    \caption{Description of the results displayed in Figure \ref{last-fig}: For each numbered curve, the corresponding  angle and number of wrinkles is given.}
\end{center}
\end{table}

The Impedance Matrix method has the advantage that it provides not only the critical compressive stretch but also the full \color{black} incremental \color{black} displacement field as obtained by solving \eqref{U}. 
Hence we have obtained the exact  \color{black} incremental \color{black}  fields shown in Figure \ref{3D} (in 3D) and in Figure \ref{last-fig} (in 2D).
Note that the incremental analysis \color{black} gives access to the amplitude of the incremental displacements up to an arbitrary scalar \color{black} factor, \color{black} and that we have chosen an exaggerated scale to display them.
\color{black}
In the examples shown in Figure \ref{last-fig}, we see that the displacements fields for the modes $k=2,3,4$ are highly localized near the surface of the compressed face, \color{black} their amplitudes \color{black} decaying rapidly with depth. 
For the mode $k=1$ (only one wrinkle), the displacements are not as localized, and propagate somewhat through the thickness of the block. This phenomenon is particularly visible in the $\Theta_0=2\pi/3$, $R_1/R_2=0.2$ case.


\section*{Acknowledgements}


Partial funding from the Royal Society of London (International Joint Project grant for MD, RWO, LV), from the Istituto Nazionale di Alta Matematica (Marie Curie COFUND Fellowship for LV; Visiting Professor Scheme for MD, IS), and from 
the Italian Ministry of Education, Universities and Research (PRIN-2009 project ``Matematica e meccanica dei sistemi biologici e dei tessuti molli"  for IS) is  gratefully acknowledged.   
We are also indebted to Jerry Murphy (Dublin City University) for helpful discussions and to Artur Gower (National University of Ireland Galway) for  technical support.


\begin{thebibliography}{99}

\color{black}
\bibitem[Aron et al.(1998)]{Aron98}
Aron, M., Christopher, C., Wang, Y., 1998. 
On the straightening of compressible, nonlinearly elastic, annular cylindrical sectors. 
Mathematics and Mechanics of Solids,
3, 131--145.
 
\bibitem[Aron(2000)]{Aron00}
Aron, M., 2000. 
Some remarks concerning a boundary-value problem in non-linear elastostatics.
Journal of Elasticity,
60, 165--172.
 
\bibitem[Aron(2005)]{Aron05}
Aron, M., 2005.
Combined axial shearing, extension, and straightening of elastic annular cylindrical sectors.
IMA Journal of Applied Mathematics, 
70,  53--63. 
\color{black}

\bibitem[Biot(1963)]{Biot63}
Biot, A.M., 1963.
Surface instability of rubber in compression.
Applied  Scientific Research 
A12, 168--182.

\bibitem[Coman \& Destrade(2008)]{CoDe08}
 Coman, C.,  Destrade, M., 2008.
Asymptotic results for bifurcations in pure bending of rubber blocks.
Quarterly Journal of Mechanics and Applied Mathematics,
61, 395--414.

\bibitem[Destrade \emph{et al.}(2009)]{DeAC09}
 Destrade, M., Ni Annaidh, A.,   Coman, C.D., 2009. 
Bending instabilities of soft biological tissues. 
International Journal of Solids and Structures, 
46, 4322--4330. 

\bibitem[Destrade \& Ogden(2005)]{DeOg05}
Destrade, M.,   Ogden, R.W., 2005. 
Surface waves in a stretched and sheared incompressible elastic material.
International Journal of Non-Linear Mechanics, 
40, 241--253.

\bibitem[Destrade \& Ogden(2010)]{DeOg10}
 Destrade, M.,  Ogden, R.W., 2010. 
On the third- and fourth-order constants of incompressible isotropic elasticity, 
Journal of the Acoustical Society of America, 
128, 3334--3343. 

\color{black}
\bibitem[Destrade \emph{et al.}(2014)]{DOSV14b}
Destrade, M.,  Ogden, R.W., Sgura, I., Vergori, L., 2014. 
Straightening: Existence, uniqueness and stability. 
Proceedings of the Royal Society of London A (to appear).
\color{black}

\bibitem[Destrade \emph{et al.}(2010)]{DeOM10}
Destrade, M.,  Murphy, J.G., Ogden, R.W., 2010. 
On deforming a sector of a circular cylindrical tube into an intact tube: Existence, uniqueness, and stability. 
International Journal of Engineering Science, 
48, 1212--1224.

\bibitem[Destrade \& Scott(2004)]{DeSc04}
Destrade, M.,  Scott, N.H., 2004. 
Surface waves in a deformed isotropic hyperelastic material subject to an isotropic internal constraint.
Wave Motion, 
40, 347--357. 

\bibitem[Ericksen(1954)]{Eric54}  
Ericksen, J.L., 1954.
Deformations possible in every isotropic, incompressible, perfectly elastic body.
ZAMP 5, 466--489.

\color{black}
\bibitem[Fu(1998)]{Fu98}
Fu, Y.B., 1998. 
Some asymptotic results concerning the buckling of a spherical shell of arbitrary thickness. 
International Journal of Non-Linear Mechanics,
33, 1111--1122.

\bibitem[Fu \& Lin(2002)]{FuLi02}
Fu, Y.B., Lin Y.P. 2002.
A WKB analysis of the buckling of an everted neo-Hookean cylindrical tube. 
Mathematics and Mechanics of Solids,
7, 483--501.
\color{black}

\bibitem[Goriely \emph{et al.}(2008)]{GoDV08}
Goriely, A.,  Vandiver, R.,  Destrade, M., 2008.
Nonlinear Euler buckling.
Proceedings of the Royal Society of London, Series A,
464, 3003--3019. 

\color{black}
\bibitem[Haughton(1999)]{Haug99}
Haughton, D.M., 1999. 
Flexure and compression of incompressible elastic plates. 
International Journal of  Engineering Science. 
37, 1693--1708.

\bibitem[Haughton(2011)]{Haug11}
Haughton, D.M., 2011.
A practical method for the evaluation of eigenfunctions from compound matrix variables in finite elastic bifurcation problems.
International Journal of Non-Linear Mechanics,
46, 795--799.
\color{black}

\bibitem[Norris \& Shuvalov(2012)]{norris12}
 Norris, A.N.,  Shuvalov, A.L., 2012.
Elastodynamics of radially inhomogeneous spherically anisotropic elastic materials in the Stroh
formalism. 
Proceedings of the Royal Society of London, Series A,
468, 467--484.

\bibitem[Ogden(1997)]{Ogde97}
 Ogden, R.W., 1997. Nonlinear Elastic Deformations. Dover, New York.

\color{black}
\bibitem[Roccabianca et al.(2011)]{RoBG11}
Roccabianca, S., Bigoni, D., Gei, M., 2011. 
Long-wavelength bifurcations and multiple neutral axes in elastic multilayers subject to finite bending. 
Journal of Mechanics of Materials and Structures, 
6, 511--52.
\color{black}

\bibitem[Schallamach(1971)]{Scha71}
Schallamach, A., 1971.
How does rubber slide?
Wear 17 (191), 301--312.

\bibitem[Shuvalov(2003)]{Shuv03}
 Shuvalov, A.L., 2003.
A sextic formalism for three-dimensional elastodynamics of cylindrically anisotropic radially inhomogeneous materials.
Proceedings of the Royal Society of London, Series A,
459, 1611--1639.

\color{black}
\bibitem[Tadmor el al.(2012)]{TaME12}
Tadmor, E.B., Miller, R.E., Elliott, R.S., 2012.
Continuum mechanics and thermodynamics. Cambridge University Press.
\color{black}

\bibitem[Truesdell \& Noll(2004)]{TrusNoll} 
Truesdell, C.,  Noll, W., 2004. 
The non-linear field theories of mechanics. Springer.

 






 


\end{thebibliography}
\end{document}